\documentclass[twocolumn,showpacs,floatfix,preprintnumbers,amsmath,amssymb]{revtex4}

\usepackage[dvips]{graphicx} 
\usepackage{txfonts}
\usepackage{dcolumn}
\usepackage{bm}

\begin{document}

\preprint{APS/123-QED}

\title{On the possibility of a maximum fundamental density \\ 
and the elimination of gravitational singularities.}

\author{Gustaf Rydbeck}
\affiliation{%
Onsala Space Observatory\\
Onsala 43497\\
Sweden
}%

%
%
\date{\today}
%
\begin{abstract}
With this note we want to point out that already in the early days of
cosmology it was 
understood that negative pressure could eliminate gravitational 
singularities in a natural way e.g. E.B. Gliner, Sov. Phys. 
JETP 22(1966)378 and 
M.A. Markov, Pis'ma Zh. Eksp. Teor. Fiz. 36, No 6, 214-216 (20 Sept. 1982).
Today, with the discovery of dark energy and the strong evidence in favor of
an inflationary start of the Big Bang, the existence of negative pressure
is widely accepted.      
In fundamental physics, phase transitions are generally thought to be 
reversible (Cf. Ellis, New Astronomy Reviews
Volume 46, Issue 11, October 2002, P. 645). 
It seems likely then that if inflation has occurred, the
process should be reversible. I.e. when the increasing density in a 
collapsing universe or star reaches a certain limit it should go through a
phase transition to a medium with an equation of state of the type
$p=\omega \rho$, where $-1< \omega <-1/3$. If this phase transition is 
fundamental, i.e. occurs for all energy densities, a collapse will always 
reach a minimum radius and bounce. If the phase transition is symmetric, 
the result will lead to oscillating universes. If however the phase 
transition is associated with an hysteresis effect, a collapsing star
may, succeeding the bounce inflate into a new  universe with a
subsequent phase transition becomes dominated by ordinary 
relativistic matter. The aim of this note is study the time development
of a model which mimics this process.
\end{abstract}
%

\maketitle
\section{Introduction}
If a theory contains singularities, it seems likely that the theory 
is inconsistent or incomplete (cf Brandenberger \cite{brandenberger}). 
The singularities which appear in 
gravitational theories result if densities become infinite. It seems
possible and natural then that at some high density $\rho=\rho_{lim}$ 
limit, all energy densities transforms to a medium with an equation of state.
\begin{eqnarray}
p=\omega \; \rho_{lim}\\
-1< \omega <-1/3 \nonumber
\end{eqnarray}
The idea is not new. It was proposed by Sakharov \cite{sakharov} and
Gliner suggested in the same year that there could be a high density limit
(cf I. Dymnikova and E. Galaktionov \cite{dymnikova}).The idea was again 
discussed by Markov \cite{markov} and further by Frolov et al. \cite {frolov}.
A similar route to eliminate singularities is to introduce high curvature
modifications in Einsteins equations (cf Mukhanov and Brandenberger 
\cite{mukhanov} and Brandenberger et al.\cite{brandenbergerII}). The 
possibility that quantum effects may cause a collapsing space to bounce as it
reaches extreme curvatures has been discussed for many years (Cf Smolin 
\cite{smolin}, Martinec \cite{martinec}, Vereshchagin \cite{vereshchagin b}, Singh et al. \cite {singh},
Vereshchagin \cite{vereshchagin c} and Easson and Brandenberger \cite{easson}).\\
It appears more or less proven that dark energy with $\omega\approx -1$ exists. 
It is thought that inflation was driven by a medium also with  $\omega\approx -1$
but with much higher density. What is not so appealing is 
the notion that inflation was preceded by by a singularity (Turner \cite{turner}).
The singularity arises naturally if it is assumed that that the inflationary 
medium is of 'false vacuum' type, since radiative excitations again becomes 
dominant as the scale factor continues to decrease as we go back in time 
or becomes dominant in a collapse phase (cf .Vereshchagin  \cite{vereshchagin a}). \\
Since media with $\omega\approx -1$ is now widely accepted,
the existence of a medium with a phase transformation as suggested above 
seems both likely and appealing. We shall therefore, as Gliner did,
hypothesize that something opposite to vacuum exists, a maximum 
density limit. 
It is often claimed that the elimination of gravitational singularities
requires a change in the the structure of General Relativity
(cf page 79 Giovanni \cite{giovanni}). 
As is shown here however, an upper density limit gets rid of
gravitational singularities in a way which leaves the equations of
general relativity unchanged, equations which by now have been
tested to quite a high accuracy (cf. e.g. van Straten et. al 
\cite{Straten}, Bertotti 
et. al \cite {Bertotti}, and Kramer et. al \cite{Kramer}). The high 
curvature modifications referred to above are of course beyond 
direct observational tests.
Since there are indications that the cosmological constant
has really been constant (cf Opher and Pelinson \cite{opher}), it seems 
that the there might as well 
be a lower limit to the total energy density of space.\\ 
Calculations based on present candidates for fundamental field theories 
such as string or M theory and quantum loop theory appear extremely complex.  
Yet the corresponding classical(i.e. non-quantum) theories, e.g. Newtons law 
of gravitation, are usually quite simple. One could therefore expect that the
"classical physics version" of the birth of our universe may be describable
in terms of a "classical" model equation of state containing a  "simple" transition 
from relativistic energy density to "inflationary like" energy-density where the 
transition is accompanied by a "hysteresis" effect (section \ref{modelsec}). \\
Since a model with a transition to "soft" inflation i.e. with an $\omega>-1$
produces to small universes the model transition is limited to case where 
$\omega$ approaches -1 for high densities.
\section{ The Friedmann equation }
In general  the pressure $P_{i}$ of the medium $i$ is related
to its density $\rho_{i}$ by the equation of state 
\begin{equation}
p_{i}=\omega \rho_{i} c^{2} 
\end{equation}
If we assume that a volume  $V$  is related to a scale factor
$a$ as $V \sim a^{3}$ and that the density is related to the volume as
\begin{equation}
\rho= \rho_{0} \left(\frac {V_{0}} {V} \right)^{n/3}= \rho_{0} \left(\frac {a_{0}} {a} \right)^{n}
\end{equation}
then $\omega=n/3-1$.
If $a$ is the scale-factor in an isotropic and homogeneous universe then the
time development of $a$ is given by the Friedmann equation
\begin{eqnarray}
\label {fried1} 
\dot{a}^{2}={\cal{G}} \rho \; a^{2} \\
\rho=\sum_ {i}\rho_{i}     \nonumber
\end{eqnarray}
where ${\cal{G}}=8\pi G/3 $ and G is the gravitational constant.
$\rho_{i}$ are the various energy densities.
The densities in the equation are the curvature 
energy density $\rho_{cu}$ , the cold matter energy density $\rho_{m}$, the 
radiative or relativistic energy density $\rho_{re}$  the
dark energy density $\rho_{de}$ and "inflationary" energy density $\rho_{ie}$.
Of these densities it is generally thought that only the curvature energy 
density can be negative. The densities may be related to a given time 
$t=t_{1}$, where $a_{1}=a(t_{1})$. For the well-known densities we have that
\begin{eqnarray}
%
\rho_{cu}=&\rho_{cu,1}(a_{1}/a)^{2} \nonumber \\
\rho_{m}=&\rho_{m,1}(a_{1}/a)^{3}\\
\rho_{re}=&\rho_{re,1}(a_{1}/a)^{4} \nonumber
\label{rho1}
\end{eqnarray}
The curvature density $\rho_{cu,1}$ is related to the curvature $\kappa_{1}$ by
\begin{equation}
\rho_{cu,1}=-\frac{c^{2} \kappa_{1} }{\cal{G}}= -\frac{c^{2} }{{\cal{G}} r_{1}^{2}}
\end{equation}
If $\rho_{cu,1}=\rho_{0}=\rho_p$, the planck density then the corresponding curvature 
radius is $r_{0}=\sqrt {\frac{3}{8\pi}}  \;  l_{p}$ where $l_{p}$ is the planck length.\\
As already mentioned (cf Opher and Belinsky\cite{opher}), recent observational results 
indicate that $\omega_{de}=-1$, i.e. that the density
\begin{equation}
\rho_{de}=\rho_{de,1}
\end{equation}
 is constant.  
The right hand side of eqn. \ref{fried1} may be considered as a negative 
gravitational potential energy. It is obvious then that if $n> 2$ or $\omega> -1/3$ 
the corresponding potential approaches minus infinity as $a\rightarrow 0$, i.e.
if the corresponding density is dominant, the universe will "fall" to a point  
singularity. If  $n< 2$ , the 
potential approaches minus infinity as $a\rightarrow \infty$, i.e. if the corresponding
 density is dominant
the universe will "fall" or inflate towards larger scales. If $n< 0$ , the energy density 
will increase with $a$. This energy is called phantom energy. One might further 
note that if $n< 2$
and the curvature energy density is negative, the scale-factor a is always positive
so a singularity cannot develop. 
\section{The de Sitter Universe with positive curvature}
The density components are in this case limited to a constant inflation driving part
$\rho_{ie}$ and a negative curvature part $\rho_{cu}$ so that the Friedmann 
equation becomes
\begin{equation} 
\dot{a}^{2}={\cal G} (\rho_{ie,0}\; a^{2}+\rho_{cu,0}\; a^{2}_{0})
\label{fried2}
\end{equation}
The solution to eqn. \ref{fried2} is
\begin{equation}
a(t)=a_{0}/2 \;\;(\exp( t/t_{0})+ \exp(-t/t_{0}) )
\end{equation}
where an arbitrary time constant is chosen so that $a(t)$ has
the minimum $a(t)$ for t=0 .
\begin{equation}
t_{0}=\left({{\cal G} \; \rho_{ie,0}}\right)^{-1/2}
\label{tzero}
\end{equation}
Further
\begin{eqnarray}
\rho_{ie,0}=\rho_{ie,1} \\ 
\rho_{ie,0}=-\rho_{cu,0} \nonumber
\end{eqnarray}
where $\rho_{cu,0}$ is the curvature density at $t=0.0$.
The radius of curvature at t=0 is 
\begin{equation}
{\cal R}=\sqrt{\frac{- c^{2}}{{\cal{G}} \; \rho_{cu,0}}}
\end{equation}
This de Sitter Universe with a finite positive curvature term has the advantage that it
does not start from a singularity but has it's origin in a previous collapse and it 
describes an inflationary origin which is how we think our universe started.  
\section{A model for the creation of a Universe from a collapsing star}\label{modelsec} 
We shall now give a simple model of the process we have discussed above. The major difference
between the previous proposals discussed in the introduction and ours is the introduction
of "friction", i.e. a term which depends 
on the expansion velocity. The term introduces true time evolution to the model, which 
without this term would just oscillate eternally. Consider a 
collapsing star where the central region is uniform and isotropic. The time evolution of
the metric of this center can then be described by the Friedmann equation (MTW 
\cite{misner}, page 852 eqn. 32.11). The iterative form of the very simple model 
equation of state for matter at high density used in our computer code is
\begin{equation}
\rho_{hd,2}=\rho_{hd,1} \left(\frac{a_{1}}{a_{2}} \right)^{4\left(1-\frac{\rho_{hd,1}}{\rho_{lim}}\right)
\left(1-\gamma \frac{v_{lim,1}}{c}\right)} 
\label{model}
\end{equation}
where $\rho_{hd}$ is the density of the relativistic high density medium and $\rho_{lim}$ 
the high density limit. $v_{lim}$ is the velocity at the event horizon distance $a_{lim}$
of the limit density. Note that the event horizon distance at the density $\rho_{hd}$ is 
larger, i.e. $\frac{v_{lim}}{c}<1$ if $\rho_{hd}<\rho_{lim}$. $a_{1}$ and $a_{2}$ are 
supposed to be differentially close, i.e. $a_{2}=a_{1}+da$ and 
$\rho_{hd,2}=\rho_{hd,1}+d\rho_{hd}$. The term 
$\left(1-\frac{\rho_{hd}(a)}{\rho_{lim}}\right)$ assures that density stops increasing as 
$\rho_{lim}$ is approached.The term $\left(1-\gamma \frac{v_{m}}{c}\right)$, where c
is the speed of light is a hysteresis 
term, i.e. an expansion velocity dependent "friction" term. This term may cause 
enormous inflation and entropy increase.
The "natural" value for the $\gamma$ constant would be one. This would cause the
inflation to be truly enormous ( the Hubble time would be tiny in comparison to the 
duration of the inflation ) so we have introduced the $\gamma$ 
parameter to limit inflation,  but $\gamma=1.0$ is still possible. Turok's string 
driven inflation (cf Turok \cite{turok}) does lead to a density limit, but does not seem to
include a velocity dependence, while intuitively I would have expected such a term.\\
We may now use the Friedmann equation \ref{fried1} to replace $\frac{v_{lim}}{c}$ in eqn. 
\ref{model} by $\pm \left(\frac{\rho_{}}{\rho_{lim}}\right)^{1/2}$, where $\pm$ is negative
under contraction and positive under expansion.
\begin{equation}
\rho_{hd,2}=\rho_{hd,1} \left(\frac{a_{1}}{a_{2}} \right)^{4\left(1-\frac{\rho_{hd,1}}{\rho_{lim}}\right)
\left(1\mp\gamma \left(\frac{\rho_{1}}{\rho_{lim}}\right)^{1/2}\right)}
\label{model1}
\end{equation}
Note that if $\rho_{hd}<<\rho_{lim}$ the usual equation of state for relativistic densities 
is obtained. The differential form of eqn. \ref{model1} is
\begin{equation}
\frac {d\rho_{hd}}{\rho_{hd}\; 4  \left(1-\frac{\rho_{hd}}{\rho_{lim}}\right)
\left(1\mp\gamma \left(\frac{\rho}{\rho_{lim}}\right)^{1/2}\right)}=-\frac{da}{a}
\label{model2}
\end{equation}
\begin{figure}
\resizebox{1.\hsize}{!} {\rotatebox{-90}{\includegraphics {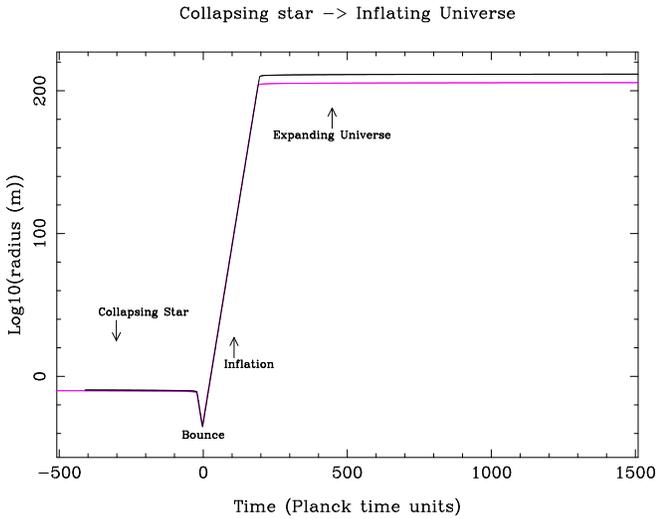}}}
\caption { Our model curvature radius at the homogeneous and isotropic and center of a star
as a function of time. As the density reaches a maximum the radius will bounce at a minimum 
value after which it starts an inflationary expansion. The initial 
curvature radius is $3 \cdot 10^{-10}$ m corresponding to a curvature density of $-1.8 \cdot 10^{45}$ kg m$^{-3}$. The
initial densities for the red and black curve are $5\times10^{87}$ and $\times10^{90}$
kg m$^{-3}$ respectively. Five hundred times greater input density at the same input curvature 
radius leads to about $10^{6}$ times greater output curvature radius, but virtually no difference
in density at later times ($> 1000$ Planck times). } 
\label{radius}
\end{figure}
\begin{figure} 
\resizebox{1.\hsize}{!} {\rotatebox{-90}{\includegraphics {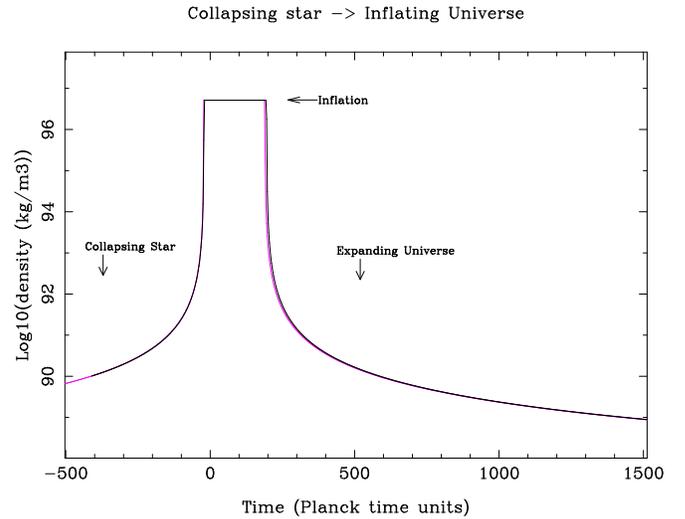}}}
\caption {The density as a 
function of time with model parameters identical to those of figure \ref{radius}.
The "sharp" corners at the beginning and end of inflation are in fact "soft" in a
magnified plot.}
\label{density}
\end{figure}
The pressure becomes
\begin{equation}
p=\left(\frac{4}{3}\left(1-\frac{\rho_{hd}}{\rho_{lim}} \right) 
\left(1\mp \gamma \frac{\rho}{\rho_{lim}}\right)-1\right) \rho_{hd} c^{2}
\end{equation}
Assuming that the limit density is the Planck density, the above equation leads 
to a cosmology described by figure\ref{radius} and \ref{density}. We have here chosen 
$\gamma=0.8$.
Our model only treats the central part of the collapsing star. The geometry and 
the time-evolution of the space away from the center where the density is not isotropic and
homogeneous may follow a complex evolution which depends on transition region constraints. 
If we assume however that the central region is sufficiently large then density or 
rarefaction waves will not have had time to reach the inner region before the expansion
velocity of
the corresponding scale-factor is larger than $c$, the speed of light. What happens outside 
this inner region then becomes unimportant to the expanding cosmological space.
It would of course be interesting to know how it connects to the center of the 
collapsed star, appearing like a black hole from the outside.\\
Although our model
concerns a special case, we think that any collapsing star or "small" universe will necessarily
convert to a contracting "near de Sitter space" which then inflates to large scales approximately
as described by the model. \\
It is interesting that for different input densities $\rho_{hd,1}$ and $\rho_{hd,2}$ at the 
same input radius, the output density becomes the same for times larger than $\sim 1000$ Planck 
times while the output radii are approximately related like 
$\frac{a_{2}}{a_{1}}=\frac{\rho_{hd,2}}{\rho_{hd,2}} $. The reason for this is that an 
increased input density
makes the universe stay slightly longer in the inflationary mode while it leaves this mode 
with $\rho_{hd}\approx \rho_{lim}$. It would thus appear that our model produces  
extreme large and very homogeneous and isotropic expanding universes.\\
Curvature fluctuations of quantum mechanical 
origin can be produced and inflated to astronomical scales in the "standard way" 
(cf e.g. Taylor and Rowan-Robinson \cite{taylor} or Lyth and Riotto\cite {lyth}).\\
\section{Discussion} 
It has been suggested that the total energy density of space may 
have an upper and a lower bound, and that the approach to the
upper limit depends on the expansion velocity. The upper bound 
and the velocity dependence would remove 
future singularities from collapsing stars, and instead lead to 
the formation of an exponentially, or probably many, expanding 
homogeneous and isotropic spaces, i.e. inflationary like 
situations. Seen from outside the collapsing star a black hole 
would still form. It is only in the microscopic center of the black 
hole that things would be different. How this inflating center connects to the 
surrounding space
would require detailed physical calculations which are beyond the scope
of this note. Our Universe could be one result from such a collapse.
Unfortunately the possibility to experimentally or observationally 
prove our suggestion seems exceedingly small.\\ 
If the initial collapsing star is small, the ensuing universe might not 
expand sufficiently for the dark energy density to overtake the negative
curvature density. The universe will then re-collapse and go through a second
bounce. The output universe will now grow to considerably larger radii so that
the dark energy term can become dominant. The end product would thus always
be a more and more empty and accelerating universe dominated by dark energy.
\\
The removal of singularities would put the problem of vanishing 
information at the center of black holes (see e.g. Giusto and 
Mathur \cite{giusto})in a different perspective. I.e. the information
is not necessarily destroyed at the center of the black hole, it might
instead be expelled into expanding "new spaces" (see Easson and Brandenberger
\cite{easson}).\\
In one particular aspect our proposition contains "new physics" and that is 
that the medium in the high density limit cannot be "excitable"  since the 
excitation energy might push the total energy density past the density limit
(Cf Vereshchagin \cite{vereshchagin a}). The medium can only be "de-excited"
and should perhaps be thought of as a large cosmological constant. 
Note that this does not imply any limits to curvature fluctuations which
would then be the sole information propagator. \\ 
Particle physics would concern excitations between an upper 
and a lower density limit. Markov \cite{markov} in fact suggests that such 
limits should be a guiding principle in the search for the fundamental field 
theory. Both string theory (cf Turok \cite{turok}) and quantum loop theory
(cf Singh \cite{singh} and Vereshchagin\cite{vereshchagin c}) do seem to 
contain upper density limits but may lead to other problems (Cailleteau et al.
\cite{cailleteau}) and are both far from settled theories. 
Will they also lead to a "friction" 
term?, the term that induces "life" to space, i.e. a space that eternally 
renews itself by the creation of new inflating universes in the centers of 
collapsing stars. \\
Many others have employed negative pressure to avoid the formation of the central
singularity in a black hole eg.  Mbonye and Kazanas \cite{mbonye}.\\ 

\section{summary}
We have pointed out that 
\begin{itemize}
\item gravitational singularities are naturally eliminated by
the introduction of a fundamental high density limit leading to a 
phase transition to a medium with negative pressure as this limit is approached.
\item instead of the formation of singularities at the center of a collapsing star
new inflating universes will form. These universes inflate to cosmological scales if a
hysteresis effect is introduced in the phase transition.\item the upper density limit and the hysteresis effect  leaves the equations 
of general relativity intact and keeps the energy-density positive, i.e.
the weak energy condition is satisfied.
\end{itemize}

%

\end{document}